\documentclass[runningheads]{llncs}
\usepackage{graphicx}
\usepackage{acronym}
\usepackage{longtable}
\usepackage{booktabs}
\usepackage{subfig}
\usepackage{multirow}
\usepackage[table]{xcolor}

\usepackage{hyperref}

\newacro{EU}[EU]{European Union}
\newacro{GDPR}[GDPR]{General Data Protection Regulation}
\newacro{ML}[ML]{machine learning}
\newacro{MLP}[MLP]{multilayer perceptron}
\newacro{AI}[AI]{artificial intelligence}
\newacro{ISO}[ISO]{International Organization for Standardization}
\newacro{IML}[iML]{interpretable machine learning}
\newacro{PCA}[PCA]{principal component analysis}
\newacro{XAI}[xAI]{explainable artificial intelligence}
\newacro{HPO}[HPO]{Hyperparameter Optimization}
\newacro{AIED}[AIED]{Artificial Intelligence in Education}
\newacro{GenAI}[GenAI]{Generative Artificial Intelligence}
\newacro{LLM}[LLM]{Large Language Model}
\newacro{AIS}[AIS]{Adaptive Instructional Systems}
\newacro{Interop}[Interop]{Interoperability}
\newacro{AISC}[AISC]{IEEE Artificial Intelligence Standards Committee}

\begin{document}
\title{Developing and Deploying Industry Standards for Artificial Intelligence in Education (AIED): Challenges, Strategies, and Future Directions}

%
\titlerunning{Industry Standards for AIED}
%


\author{Richard Tong\inst{1,2} \and
Haoyang Li\inst{3} \and
Joleen Liang\inst{3} \and
Qingsong Wen\inst{2}}
\authorrunning{R. Tong et al.}
%
\institute{IEEE Artificial Intelligent Standards Committee, IEEE, USA\and
Squirrel Ai Learning, USA\\
\and
Squirrel Ai Learning, China\\
\email{richard.tong@ieee.org,\{derekli,joleenliang,qingsongwen\}@squirrelai.com}}

\maketitle              

\begin{abstract}
The adoption of \ac{AIED} holds the promise of revolutionizing educational practices by offering personalized learning experiences, automating administrative and pedagogical tasks, and reducing the cost of content creation. However, the lack of standardized practices in the development and deployment of AIED solutions has led to fragmented ecosystems, which presents challenges in interoperability, scalability, and ethical governance. This article aims to address the critical need to develop and implement industry standards in AIED, offering a comprehensive analysis of the current landscape, challenges, and strategic approaches to overcome these obstacles. We begin by examining the various applications of AIED in various educational settings and identify key areas lacking in standardization, including system interoperability, ontology mapping, data integration, evaluation, and ethical governance. Drawing on insights from stakeholders across the educational technology spectrum, we propose a multi-tiered framework for establishing robust industry standards for AIED. This framework emphasizes collaborative efforts among educational institutions, technology providers, regulatory bodies, and learners to ensure that AIED solutions are accessible, equitable, and effective for diverse populations of learners, educational settings, and socioeconomic environments. In addition, we discuss methodologies for the iterative development and deployment of standards, incorporating feedback loops from real-world applications to refine and adapt standards over time. The paper also highlights the role of emerging technologies and pedagogical theories in shaping future standards for AIED. Finally, we outline a strategic roadmap for stakeholders to implement these standards, fostering a cohesive and ethical AIED ecosystem. By establishing comprehensive industry standards, such as those by \ac{AISC} and \ac{ISO}, we can accelerate and scale AIED solutions to improve educational outcomes, ensuring that technological advances align with the principles of inclusivity, fairness, and educational excellence.


\end{abstract}
\section{Introduction}

Artificial Intelligence has emerged as a transformative force in the educational landscape, promising to revolutionize teaching and learning practices by offering personalized experiences, automating tasks, and providing data-driven insights~\cite{li2024bringing}. One key enabler for the application of \ac{AIED} is the establishment of standards and regulations for practitioners to develop simple, scalable, safe and trustworthy AI systems.\cite{tonglee202305} Despite several efforts by different research institutions and government organizations, this innovative landscape is currently marred by a significant challenge: the absence of standardized practices in the development and deployment of AIED solutions. This lack leads to fragmented ecosystems, which pose substantial hurdles in achieving interoperability, scalability, and ethical governance.

This paper aims to address the critical need for developing and deploying industry standards in AIED. We provide a comprehensive analysis of the current landscape, challenges, and strategic approaches to overcome these hurdles. Drawing on insights from stakeholders across the educational technology spectrum, we propose a multi-tiered framework for leveraging, building, and extending industry standards, especially the IEEE Learning Technology Standards and IEEE Artificial Intelligence Standards, for AIED practices and applications.

The paper is structured as follows: Section 1 introduces the background, the purpose, the challenges and the general approach of AIED standardization. Section 2 examines the landscape of diverse applications of AIED across various educational settings and identifies key areas lacking standardization. Section 3 presents our proposed multi-tiered framework for developing and deploying AIED standards, emphasizing collaborative efforts among stakeholders. Section 4 discusses methodologies for iterative development and deployment of standards. Section 5 highlights the role of emerging technologies and pedagogical theories in shaping future AIED standards. Finally, Section 6 outlines a strategic roadmap for stakeholders to implement these standards.

\section{Applications and Standard Landscape for AIED}

AIED has found applications across a wide range of educational settings, from K-12 classrooms to higher education institutions and corporate training programs. Some key areas where AIED is being leveraged include:

\begin{itemize}
    \item Intelligent Tutoring Systems (ITS): ITS use AI algorithms to provide personalized instruction, feedback, and guidance to learners based on their individual needs and progress.
    \item Adaptive Learning Platforms: These platforms dynamically adjust the content, pace, and difficulty of learning materials based on learner performance and preferences.
    \item Learning Analytics and Educational Data Mining: AI techniques are used to analyze large datasets of student interactions and performance to derive insights and recommendations for improving teaching and learning outcomes.
    \item Automated Assessment and Grading: AI algorithms can automate the grading of essays, short answers, and programming assignments, reducing the burden on educators and providing timely feedback to learners.
    \item Chatbots and Virtual Assistants: AI-powered conversational agents can provide 24/7 support to learners, answering queries, offering guidance, and facilitating learning interactions.
\end{itemize}

Despite the growing adoption of AIED, the lack of standardization poses significant challenges:

\begin{enumerate}
    \item Interoperability: The absence of common data models and communication protocols hinders the seamless integration and exchange of data between different AIED systems and platforms.
    \item Scalability: Without standardized approaches for designing, developing, and deploying AIED solutions, it becomes difficult to scale these technologies across diverse educational contexts and learner populations.
    \item Ethical Governance: The lack of standardized ethical guidelines and principles for AIED raises concerns about privacy, fairness, transparency, and accountability in the use of these technologies.
\end{enumerate}

\section{A Multi-Tiered Framework for AIED Standards}

To address the challenges and standardization needs in AIED, we propose a multi-tiered framework that leverages, builds upon, and extends existing industry standards, especially the IEEE Learning Technology Standards and IEEE Artificial Intelligence Standards. This framework emphasizes collaborative efforts among key stakeholders, including educational institutions, technology providers, regulatory bodies, practitioners, educators, and learners.

\subsection{The proposed framework consists of three tiers:}

\subsubsection{Tier 1: Leveraging Existing Standards}

The first tier focuses on identifying and leveraging existing industry standards relevant to AIED. This includes standards such as:

\begin{enumerate}
    \item IEEE Learning Object Metadata (LOM): A standard for describing learning resources and enabling their discovery, sharing, and reuse across different systems.
    \item Experience API (xAPI): A specification for capturing and sharing learning experiences across diverse platforms and contexts.
    \item IMS Global Learning Consortium Standards: A suite of standards for enabling interoperability, accessibility, and innovation in educational technology, including the Learning Tools Interoperability (LTI) standard for integrating learning applications with learning management systems.
    \item IEEE and ISO Artificial Intelligence Standards: Emerging standards for ensuring the ethical, transparent, and accountable development and deployment of AI systems, such as the IEEE P3396, ISO42001 series of standards~\cite{ieee19}.
\end{enumerate}
By adopting and aligning with these existing standards, AIED solutions can achieve greater interoperability, scalability, and compatibility with the broader educational technology ecosystem.

\subsubsection{Tier 2: Building AIED-Specific Standards}

The second tier involves developing new standards, especially IEEE P3428, specifically tailored to the unique requirements and challenges of AIED. These standards should address the four key areas identified earlier: domain model, pedagogical model, learner model, and interaction model following the work previously conducted on IEEE P2247.x and around Self-improvable Adaptive Instructional Systems~\cite{Tong2020Architecture}.

\begin{enumerate}
    \item Domain Model Standards: Develop standards for representing knowledge domains, learning objectives, and instructional content in a structured and machine-readable format. This may involve extending existing standards like the IEEE Learning Object Metadata (LOM) to incorporate AIED-specific attributes and relationships.
    \item Pedagogical Model Standards: Establish standards for representing pedagogical strategies, instructional design principles, and best practices in AIED. This may include defining a common vocabulary and taxonomy for describing pedagogical approaches, as well as specifying guidelines for designing effective AIED interventions.
    \item Learner Model Standards: Create standards for representing learner profiles, preferences, and performance data in a consistent and interoperable manner. This may involve defining a common data model for learner attributes, as well as specifying protocols for securely sharing and exchanging learner data across different AIED platforms.
    \item Interaction Model Standards: Develop standards for designing and implementing user interfaces, interaction patterns, and communication protocols in AIED. This may include defining best practices for designing intuitive and accessible interfaces, as well as specifying standards for integrating AIED components with existing learning management systems and educational platforms.
\end{enumerate}
The development of these AIED-specific standards should be a collaborative effort involving diverse stakeholders, including researchers, practitioners, educators, and technology providers. This collaborative approach ensures that the standards are grounded in real-world needs and experiences, and that they are widely adopted and supported by the AIED community.

\subsubsection{Tier 3: Extending Standards for Emerging Technologies}

The third tier of the framework focuses on extending existing and newly developed AIED standards to accommodate emerging technologies such as generative AI, large language models (LLMs), and advanced pedagogical theories.

\begin{enumerate}
    \item Generative AI Standards: Develop standards for leveraging generative AI techniques, such as machine learning-based content creation and personalization, within AIED systems. This may involve guidelines for training and deploying generative models, as well as specifying protocols for integrating generated content with existing AIED components.
    \item LLM Standards: Establish standards for incorporating LLMs, such as GPT and Claude, into AIED solutions. This may include defining interoperability among LLM agents (IEEE P3394), defining best practices for fine-tuning LLMs for educational purposes, as well as specifying guidelines for ensuring the ethical and responsible use of LLMs in educational contexts (IEEE P3395).
    \item Pedagogical Theory Standards: Extend AIED standards to incorporate advances in pedagogical theories, such as constructivism, social learning, and self-regulated learning. This may involve defining guidelines for designing AIED interventions that align with these theories, as well as specifying protocols for evaluating the effectiveness of theory-informed AIED solutions.
\end{enumerate}
By extending AIED standards to accommodate emerging technologies and pedagogical theories, the framework ensures that the standards remain relevant and adaptable to the rapidly evolving landscape of educational technology.

\subsection{Key Standards for AIED Adoption and Integration}

To take the 3-tiered approach to leverage AIED related standards to build solutions, here are some key IEEE and non-IEEE standards:

\begin{figure}
    \includegraphics[width=1.0\textwidth]{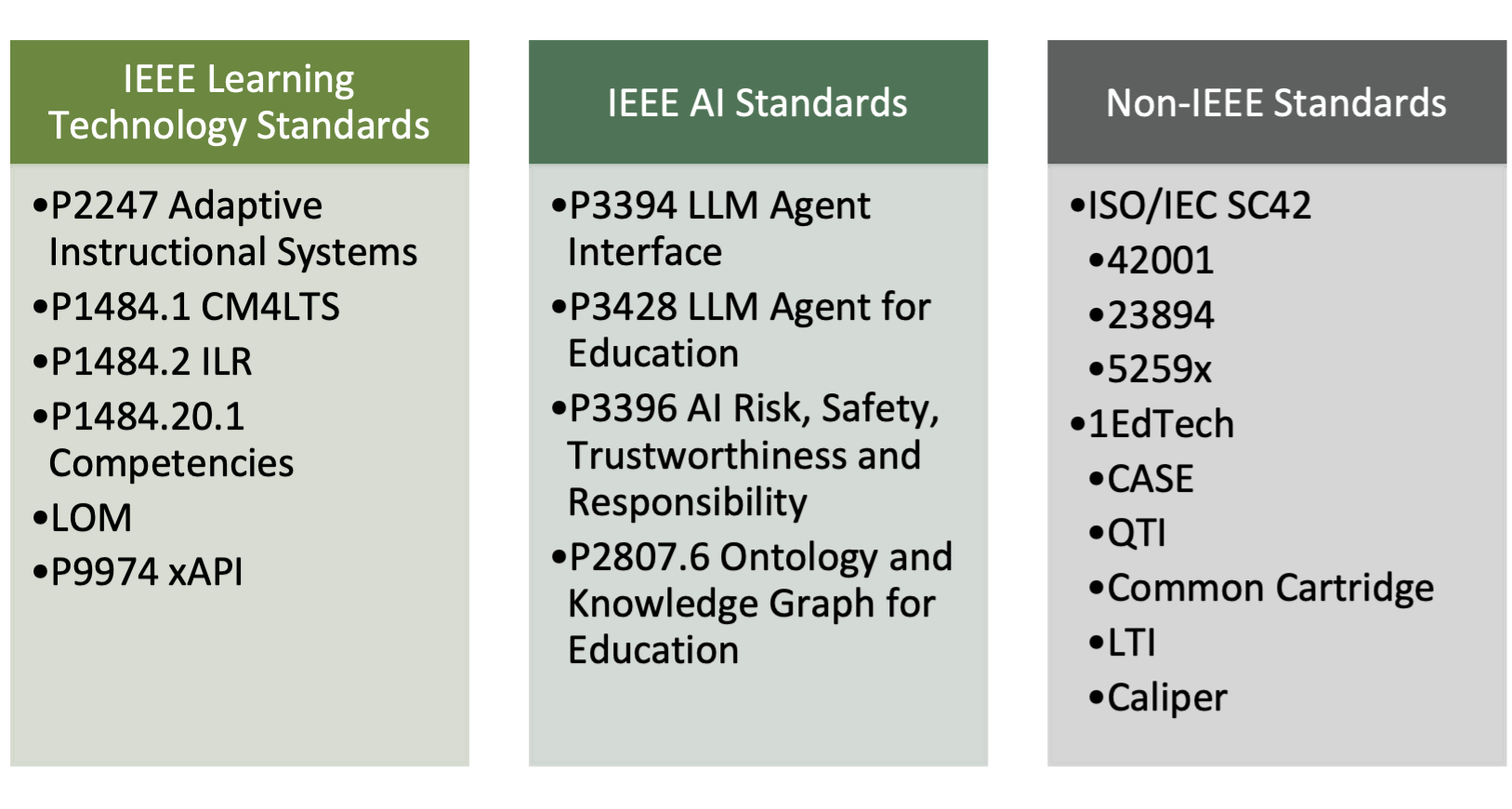}
    \caption{Core AIED Standards to enable both system integration and ethical adoption of AI}
\end{figure}

\subsubsection{IEEE P3394 - Large Language Model Agent Interface}
This standard outlines protocols and interfaces for large language model agents overall. It focuses on enhancing the interoperability, security, and efficiency of these models when interacting with human agents and other AI agents as well as traditional systems.

\subsubsection{IEEE P3428 - LLM Agent for Education}
This standard, which is in its initial phases, targets the interoperability of LLM Agents developed for educational systems such as AIS. It offers guidelines and specifications for the integration and evaluation of both open-source and closed-source education-specific LLM and FM agents.

\subsubsection{IEEE P2247.2 - Adaptive Instructional System Interoperability}
This standard is crucial for the interoperability of adaptive instructional systems. It provides guidelines for the vertical and horizontal integration of AIS, ensuring a cohesive and effective learning experience and ease of integration.

\subsubsection{IEEE P3396 - AI Risk, Trustworthiness, Safety, and Responsibility}
This recommended practice provides a thorough framework for understanding and defining AI risks, safety, trustworthiness, and responsibility. It aims to manage these aspects while maintaining the benefits of innovation. The practice considers the global context and promotes responsible AI adoption, governance, and collaboration. It offers a principles-based framework, examining AI's role in information generation, decision-making, and human agency. It also considers responsibilities related to AI usage throughout the entire life cycle of AI application development, deployment, and operation along with ISO23894~\cite{ISO23894}.

\subsubsection{ISO/IEC 42001 - AI management system standard}
ISO/IEC 42001:2023~\cite{ISO42001} provides a foundational framework for the responsible management of AI, addressing the challenges and ethical implications inherent in AI technologies and ensuring that AI systems are governed and managed in a way that is beneficial, ethical, and transparent across various industries~\cite{iso22989}. 

\subsection{Summary}

we can leverage international standards such as those by the IEEE Artificial Intelligence Standards Committee, IEEE LTSC, and ISO/IEC SC42 to accelerate the agent implementation for educational systems. 

\section{Methodologies for Standards Development and Deployment}

Developing and deploying industry standards for AIED requires a systematic and iterative approach that engages diverse stakeholders and ensures the standards are robust, practical, and widely adopted. We propose the following methodologies for standards development and deployment:

\begin{enumerate}
    \item Stakeholder Engagement: Involve a diverse range of stakeholders, including researchers, practitioners, educators, learners, technology providers, and policymakers, in the standards development process. This can be achieved through workshops, surveys, focus groups, and online collaboration platforms.
    \item Iterative Development: Adopt an iterative approach to standards development, starting with the identification of key requirements and use cases, followed by the drafting of initial specifications, and then refining the standards based on feedback and pilot implementations.
    \item Pilot Implementations: Encourage early adoption and pilot implementations of the proposed standards by educational institutions, technology providers, and research groups. These pilot implementations provide valuable feedback and insights for refining the standards and ensuring their practicality and effectiveness.
    \item Conformance Testing and Certification: Establish mechanisms for conformance testing and certification to ensure that AIED solutions comply with the developed standards. This may involve creating test suites, validation tools, and certification programs that assess the interoperability, scalability, and ethical compliance of AIED systems.
    \item Continuous Improvement: Regularly review and update the standards based on feedback from stakeholders, advances in technology and pedagogy, and evolving educational needs. This continuous improvement process ensures that the standards remain relevant and effective over time.
    \item Capacity Building and Training: Provide training and capacity building programs to help stakeholders understand and implement the standards effectively. This may include workshops, online courses, and mentorship programs that empower educators, researchers, and technology providers to adopt and leverage the standards in their work.
    \item Policy and Governance Frameworks: Develop policy and governance frameworks that support the adoption and enforcement of AIED standards. This may involve collaboration with regulatory bodies, educational authorities, and industry associations to establish guidelines, incentives, and accountability mechanisms for standards compliance.
\end{enumerate}
By following these methodologies, the development and deployment of AIED standards can be a collaborative, iterative, and sustainable process that engages diverse stakeholders and ensures the standards are robust, practical, and widely adopted.

\section{The Role of Emerging Technologies and Pedagogical Theories}

Emerging technologies and pedagogical theories play a crucial role in shaping the future of AIED and informing the development of industry standards. Two key areas that are poised to have a significant impact on AIED standards are generative AI and large language models (LLMs).

Generative AI techniques, such as machine learning-based content creation and personalization, have the potential to revolutionize AIED by enabling the automated generation of high-quality educational content tailored to individual learners' needs and preferences. For example, generative models can be used to create personalized learning materials, assessments, and feedback based on learners' performance data and learning objectives. To harness the full potential of generative AI in AIED, standards need to be developed for representing educational content in a structured and machine-readable format, as well as for specifying protocols for integrating generated content with existing AIED components.

LLMs, such as GPT, Gemini, and Claude, have the potential to transform AIED by enabling more natural and engaging interactions between learners and AI-powered educational systems. LLMs can be leveraged to develop intelligent tutoring systems, chatbots, and virtual assistants that can provide personalized support, answer questions, and facilitate learning discussions. However, the use of LLMs in AIED also raises ethical concerns related to bias, transparency, and accountability. To ensure the responsible and effective use of LLMs in AIED, standards need to be established for fine-tuning LLMs for educational purposes, as well as for specifying guidelines for ensuring the ethical and transparent deployment of LLMs in educational contexts.

In addition to emerging technologies, advances in pedagogical theories also have significant implications for AIED standards. Pedagogical theories, such as constructivism, social learning, and self-regulated learning, provide insights into how learners construct knowledge, interact with peers and teachers, and regulate their own learning processes. These theories can inform the design of AIED interventions that are more effective, engaging, and learner-centered. For example, standards can be developed for designing AIED systems that support collaborative learning, scaffolding, and metacognitive strategies aligned with these pedagogical theories. Moreover, standards can be extended to specify protocols for evaluating the effectiveness of theory-informed AIED solutions and ensuring their alignment with evidence-based practices.

To fully leverage the potential of emerging technologies and pedagogical theories in AIED standards, it is essential to foster interdisciplinary collaboration among researchers, practitioners, and technology providers. This collaboration can facilitate the exchange of knowledge, experiences, and best practices across different domains and ensure that the standards are grounded in cutting-edge research and real-world needs. Moreover, the standards development process should be flexible and adaptable to accommodate the rapid pace of technological and pedagogical innovations, allowing for the continuous refinement and extension of standards as new technologies and theories emerge.

\section{Strategic Roadmap for Standards Implementation}

\subsection{Roadmap adopted by IEEE AISC}
Implementing industry standards for AIED requires a strategic and coordinated effort from all stakeholders involved. We (IEEE AISC, IEEE LTSC and IEEE P3428 LLM Agent for Education WG) have jointly developed the following strategic roadmap for standards implementation and collaboration:

\begin{enumerate}
    \item Establish a Standards coalition composed of representatives from IEEE, educational institutions, technology providers, regulatory agencies, and other relevant stakeholders.
    \item Conduct Needs Assessment and Gap Analysis: Engage stakeholders in a comprehensive needs assessment and gap analysis to identify the most pressing standardization requirements and priorities in AIED. 
    \item Develop a Standards Roadmap: Based on the needs assessment and gap analysis, we have developed a detailed roadmap for standards development and deployment. 
    \item Foster Stakeholder Collaboration: Establish mechanisms for ongoing collaboration and engagement among stakeholders throughout the standards development and deployment process. 
    \item Pilot and Refine Standards: Encourage early adoption and pilot implementations of the proposed standards by educational institutions, technology providers, and research groups. Use the feedback and insights from these pilot implementations to refine and improve the standards iteratively.
    \item Develop Conformance Testing and Certification Programs: Establish robust conformance testing and certification programs to ensure that AIED solutions comply with the developed standards. These programs should assess the interoperability, scalability, and ethical compliance of AIED systems and provide clear guidelines for technology providers and adopters.
    \item Provide Training and Support: Offer comprehensive training and support programs to help stakeholders understand and implement the standards effectively. This may include workshops, online courses, and mentorship programs that empower educators, researchers, and technology providers to adopt and leverage the standards in their work.
    \item Monitor and Evaluate Standards Adoption: Continuously monitor and evaluate the adoption and impact of AIED standards across different educational contexts. Use this data to identify areas for improvement, address challenges, and measure the effectiveness of the standardization efforts.
    \item Engage in Continuous Improvement: Regularly review and update the standards based on feedback from stakeholders, advances in technology and pedagogy, and evolving educational needs. This continuous improvement process ensures that the standards remain relevant, effective, and responsive to the changing landscape of AIED.
    \item Advocate for Policy and Regulatory Support: Collaborate with policymakers, regulatory bodies, and educational authorities to advocate for policies and regulations that support the adoption and enforcement of AIED standards. This may involve establishing incentives, funding mechanisms, and accountability frameworks that encourage standards compliance and promote the responsible and effective use of AIED in education.
\end{enumerate}

\subsection{Case Notes on working with ISO and EU AI Act for AIED Policy Standardization}
Working with ISO and addressing the European Union for AIED policy standardization presents unique challenges and opportunities and the recent effort has given us unique perspective in planning and implementing the roadmap. The evolving European AI Act highlights the critical role of standards in ensuring AI systems in education are developed and implemented responsibly.

\subsubsection{Aligning with the European AI Act} 
 For AIED, it's paramount to align with the European AI Act's requirements~\cite{AiAct}. This involves adhering to established standards for risk management, data governance, and transparency, among others. It's about embedding these requirements into the AI systems' lifecycle—from design to deployment—ensuring they are not only compliant but also ethical and effective in educational settings. In general, the \ac{AI} act distinguishes three risk classes for \ac{AI}: forbidden applications, high-risk applications and uncritical applications. Whether an application falls into the high-risk category is risk-dependent, and a few high-risk applications are named in the \ac{AI} Act's appendix. Fortunately and unfortunately, Education falls into this category while military use is not considered as high risk. 

\subsubsection{Collaboration with ISO}
ISO's role in setting international standards is invaluable. The ISO/IEC JTC 1/SC 42 is the technical committee within the \ac{ISO} that is responsible for developing standards for \ac{AI}. An overview can be found on the web page of the committee.\footnote{\href{https://www.iso.org/committee/6794475/x/catalogue/p/1/u/1/w/0/d/0}{Standards by ISO/IEC JTC 1/SC 42: Artificial intelligence}} Collaborating with ISO ensures that the standards developed for AIED are globally recognized and can be adopted universally. This partnership can facilitate the creation of standards that address specific needs, such as robustness and cybersecurity, making AI applications in education safer and more reliable.

\subsubsection{Bridging Policy and Practice}
Bridging the gap between policy (EU regulations) and practice (ISO standards) is crucial. This involves translating the legal requirements of the AI Act into technical standards that can be practically implemented. It requires a deep understanding of both the regulatory environment and the technical challenges involved in AIED.

\subsection{Summary}

By following this strategic roadmap, stakeholders can work together to develop, deploy, and maintain robust and effective industry standards for AIED. These standards will not only address the current challenges related to interoperability, scalability, and ethical governance but also pave the way for a more inclusive, equitable, and innovative future of education powered by artificial intelligence.

\section{Conclusion}

The development and deployment of industry standards for AIED are crucial for harnessing the full potential of artificial intelligence in transforming education. By addressing the challenges of interoperability, scalability, and ethical governance, these standards can enable the creation of AIED solutions that are more effective, accessible, and equitable for diverse learner populations.

The proposed multi-tiered framework, which leverages existing standards, builds AIED-specific standards, and extends standards for emerging technologies, provides a comprehensive approach to standardization that engages diverse stakeholders and ensures the standards are robust, practical, and widely adopted. The methodologies for standards development and deployment, including stakeholder engagement, iterative development, pilot implementations, and continuous improvement, offer a systematic and sustainable process for creating and maintaining effective AIED standards.

Moreover, the strategic roadmap for standards implementation, which encompasses the establishment of a coalition, needs assessment, stakeholder collaboration, GenAI oriented protocol and standard development, and policy advocacy, provides a clear path forward for all stakeholders to work together in realizing the vision of standardized and responsible AIED.

\bibliographystyle{splncs04}
\bibliography{ref-ai-std,ref-regulate,reference}
%




\end{document}